\documentclass[%
 reprint,
superscriptaddress,
 amsmath,amssymb,
 aps,prl
]{revtex4-1}

\usepackage{graphicx}
\usepackage{dcolumn}
\usepackage{bm}
\usepackage{color}

\begin{document}
\preprint{APS/123-QED}
\title{Feedback Cooling of a Single Neutral Atom}
\author{Markus Koch}
\email{markus.koch@mpq.mpg.de}
\author{Christian Sames}
\author{Alexander Kubanek}
\author{Matthias Apel}
\author{Maximilian Balbach}
\author{Alexei Ourjoumtsev}
\altaffiliation[Present address: ]{Laboratoire Charles Fabry, Institut d'Optique, CNRS, Univ. Paris-Sud, Campus Polytechnique, RD 128, 91127 Palaiseau cedex, France }
\affiliation{Max-Planck-Institut f\"ur Quantenoptik, Hans-Kopfermann-Strasse 1, 85748 Garching, Germany}
\author{Pepijn W.H. Pinkse}
\affiliation{MESA+ Institute for Nanotechnology, University of Twente, 7500 AE Enschede, The Netherlands}
\author{Gerhard Rempe}
\affiliation{Max-Planck-Institut f\"ur Quantenoptik, Hans-Kopfermann-Strasse 1, 85748 Garching, Germany}
\date{\today}

\begin{abstract}
We demonstrate feedback cooling of the motion of a single rubidium atom trapped in a high-finesse optical resonator to a temperature of about 160 $\mu$K. Time-dependent transmission and intensity-correlation measurements prove the reduction of the atomic position uncertainty. The feedback increases the 1/e storage time into the 1 s regime, 30 times longer than without feedback. Feedback cooling therefore rivals state-of-the-art laser cooling, but with the advantages that it requires less optical access and exhibits less optical pumping.
\end{abstract}
\maketitle

Achieving strong coupling of a single atom to a light mode opens up new avenues for many applications ranging from sensitive detection to quantum information. Towards these goals, a variety of optical resonators such as micrometer-sized Fabry-Perot cavities \cite{Hood00}, fiber-based cavities on atom chips \cite{gehr10}, bottle resonators \cite{Poellinger09} or micro-toroidal resonators \cite{Aoki06} have been developed. They all have in common that the atom resides inside a cavity with very small mode volume and, hence, restricted optical access. Moreover, for most atomic isotopes, three-dimensional (3D) laser cooling populates different Zeeman states of the atom with a reduced coupling to the cavity mode. Both constraints render 3D laser cooling impracticable if not impossible. 

A solution is feedback cooling. The idea is to observe the moving atom and then, using a fast feedback loop, engineer a suitable dissipative force that slows down the particle \cite{Mancini00,Steck04,Lynn05}. The performance is mainly limited by the accuracy with which the atomic trajectory can be measured. So far, feedback cooling has only been implemented for charged particles like antiprotons under the name of stochastic cooling \cite{Meer85}, for electrons in Penning traps \cite{Urso03}, for ions in Paul traps \cite{Bushev06} and for micro-mechanical resonators \cite{Poggio07}. A modest increase of the storage time of a single trapped atom has been achieved by means of feedback \cite{Fischer02,Kubanek09} but cooling has not yet been demonstrated directly.

Here we report on measurements performed in a new experimental setup optimized for the demands of feedback cooling. Compared to \cite{Kubanek09}, more than 4 times higher photon detection efficiencies and faster feedback logic were implemented. As a result, we achieve average storage times exceeding 1 s. We deduce a temperature of about 160 $\mu$K and observe the improved localization of the atom. We also show that the performance of feedback cooling depends strongly on the amount of available information. Our results demonstrate the versatility of feedback cooling for all kinds of strong-coupling experiments with single atoms where 3D laser cooling is difficult to realize due to the limited optical access. 

Our $^{85}$Rb atoms are trapped inside a Fabry-Perot cavity with two spherical mirrors of different radii of curvature (R$_{1}$ = 200 mm, R$_{2}$ = 10 mm) and transmission coefficients ($T_{1}$ = 2 ppm, $T_{2}$ = 16 ppm, $L_{1}+L_{2} \approx$ 11 ppm), resulting in a finesse of F $\approx 2 \times 10^{5}$. The cavity has a length of 260 $\mu$m, yielding a cavity field decay rate of $\kappa=2\pi\times$1.5 MHz and maximum atom-field coupling constant of $g_{0}$=$2\pi\times$16 MHz. With an atomic dipole decay rate $\gamma=2\pi\times$3 MHz, this puts our experiment in the strong-coupling regime of cavity quantum electrodynamics (QED). 

Two circularly polarized lasers at 785 nm and 780 nm are simultaneously coupled into the cavity through the mirror with the lower transmission. They are on resonance with two $TEM_{00}$ modes of the cavity separated by four free spectral ranges. The laser at 785 nm stabilizes the cavity length, traps the atom, and actuates the atomic motion. The other laser at 780 nm, called the probe laser, is blue detuned by 40 MHz from the $5S_{1/2},F=3$ to $5P_{3/2}, F=4$ transition. It is $\sigma^{+}$ polarized, driving the closed $m_{F}=3$ to $m_{F}=4$ cycling transition. The detuning between the atom and the cavity leads to cavity cooling along the cavity axis \cite{Maunz04}. Under these conditions the trapping times are limited by momentum diffusion transverse to the cavity axis. The light transmitted through the cavity is separated spectrally using interference filters. The probe light is detected with single-photon counting modules, whose clicks are recorded by a home-made FPGA-based (field-programmable gate array) photon counter that also activates the feedback algorithm. Single atoms are loaded into the cavity by means of a pulsed atomic fountain. Upon the arrival of an atom, which is heralded by a drop of the probe-light transmission, the power of the dipole laser is increased, trapping the atom. 

The probe-light transmission increases with the distance of the atom to the cavity axis, i.e., for decreasing coupling. The transmission during two intervals with a duration of 13 $\mu$s (the integration time) is evaluated and compared. If the transmission during the earlier interval surpasses the transmission during the latter interval by a certain threshold (typically 3 photon clicks), i.e. if the transmission is decreasing with time corresponding to an atom moving inwards, the dipole trap is switched low, to about 400 $\mu$K, to minimize the kinetic energy gained by the atom. If, however, the transmission difference is below that threshold, the trap depth is switched high, typically to 950 $\mu$K, increasing the potential gradient for the outgoing atom \cite{Kubanek09}. The integration time, threshold and dipole powers were optimized for good feedback performance. 

\begin{figure}
\includegraphics[width=8cm]{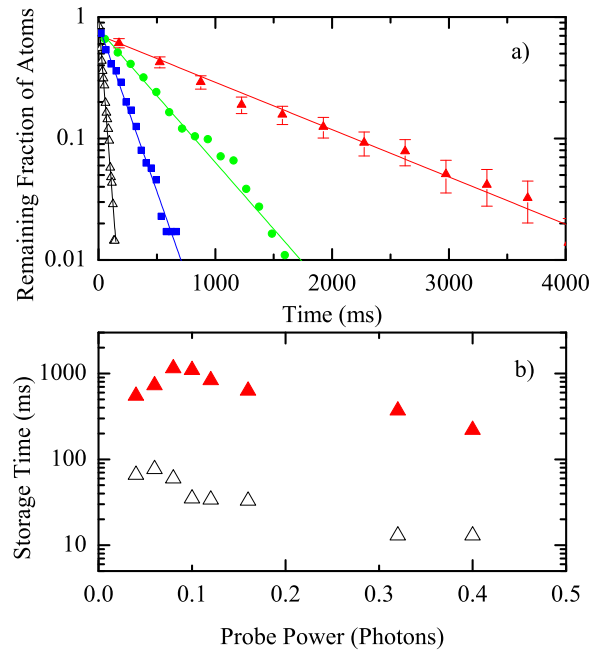}
\caption{\label{Figure1}(Color online) (a) Evolution of the probability of an atom to remain trapped without feedback ($\vartriangle$) and with feedback for 25 $\%$ ($\blacksquare$), 50 $\%$ ($\bullet$) and 100 $\%$ ($\blacktriangle$) signal. The 1/e storage times are 35 ms, 160 ms, 400 ms and 1100 ms, respectively. For clarity, error bars are only shown for one curve. (b) Storage time with ($\blacktriangle$) and without ($\vartriangle$) feedback versus probe power in units of empty cavity photon number. The error bars obtained from the fit are smaller than the symbols.}
\end{figure}

High photon detection efficiencies are crucial for an accurate position measurement. Compared to our earlier work \cite{Kubanek09}, the detection efficiency has been enhanced by more than a factor of 4 (from about 5\% to 23\%), due to the use of an asymmetric cavity and an improved detection setup. The better signal-to-noise ratio enables a more reliable position measurement at lower probe powers where less heating occurs. The significance of high detection efficiencies is highlighted in Fig. \ref{Figure1} a, which shows a storage time measurement without feedback and with feedback for different attenuations (0 $\%$, 50 $\%$ and 75 $\%$) of the transmitted probe light. To compensate for off-resonant scattering to the $F=2$ hyperfine ground state, which occurs at a small rate of about 12 Hz due to nonperfect circular polarizations and residual magnetic fields, we employ a weak repumping laser perpendicular to the cavity axis. This laser is not mandatory for feedback cooling but allows us to observe storage times exceeding 80 ms. The storage time of 200 individual atoms is evaluated for each measurement and the fraction of atoms that has remained trapped until a time $t$ is calculated. We fit an exponential $p(t)=Ae^{-t/\tau}$ to these curves with the amplitude $A$ and the $1/e$ storage time $\tau$ as fit parameters. Some atoms (about 10$\%$-30$\%$) leave the trap within the first time bin thereby reducing the amplitude to 0.7-0.9, likely because they are not trapped at all or they are trapped at a position where the trap and probe modes do not overlap and the feedback does not work properly. We observe traces with durations of up to 12 s and a $1/e$ storage time of 1100 ms, compared to 35 ms without feedback. As we attenuate the transmitted probe light, the storage time decreases in proportion to the attenuation, demonstrating the importance of high detection efficiency. 

This observation seems to indicate that an even better performance can be achieved for larger probe power. However, this leads to more heating of the atom as shown in Fig. \ref{Figure1} b. Here, the $1/e$ storage time is plotted versus the impinging probe power with and without feedback. By applying feedback we observe an increase of the storage time by more than 1 order of magnitude over a broad range of powers with a peak at a probe power of 0.1 photons in the empty cavity (without the atom), where the improvement is 30-fold. For higher probe powers the absolute storage time decreases as the signal increase does not fully compensate for the increase in heating. Also, for lower probe powers one sees a decrease of the performance of the feedback cooling, for two reasons. First, the signal becomes weaker thus degrading the accuracy of the velocity estimation. Second, as can be seen from the decrease of the storage time without feedback, for very low probe powers cavity cooling along the cavity axis can no longer compensate for axial parametric heating caused by intensity fluctuations of the dipole trap and other axial heating mechanisms \cite{Maunz04}. This opens up another loss channel along the cavity axis that is not counteracted by the feedback cooling. 

\begin{figure}
\includegraphics[width=8cm]{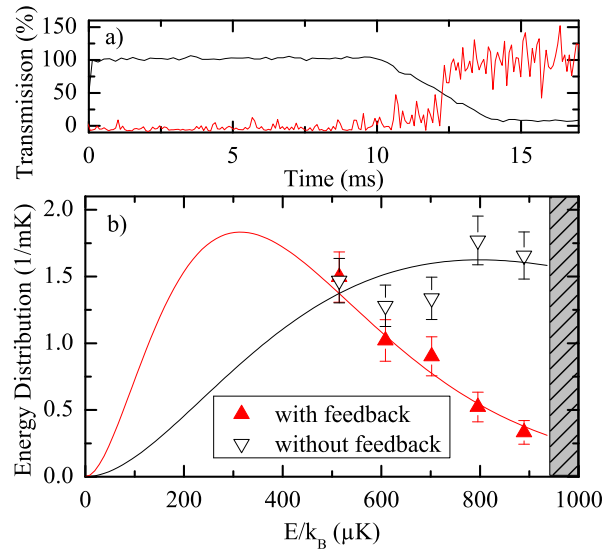}
\caption{\label{Figure2}(Color online) (a) Typical transmission signal (red) from an atom that escapes while the dipole trap (black) is ramped down. (b) Reconstructed energy distribution of the atom. Fitting a Boltzmann distribution yields a temperature of 160$\pm$5 $\mu$K with feedback and 400$\pm$50 $\mu$K without feedback.}
\end{figure}

We use a similar technique as in Ref. \cite{Alt03} to obtain a temperature estimate of the atom. For this purpose the potential depth is lowered linearly from 950 $\mu$K to 100 $\mu$K within 4 ms. From the measured escape time of the atom we infer the potential depth and therefore the energy of the atom (cf. Fig. \ref{Figure2} a). Following \cite{Alt03} we take into account adiabatic cooling by assuming that the action of the atom $S(E,U)=4\int^{x_{max}}_{0}2m\sqrt{E-V(x,U)}dx $ is conserved. Here E designates the energy, $x_{max}$ the oscillation amplitude and $V(x,U)=-U e^{-x^2}$ a Gaussian potential with depth $U$. Solving numerically the equation $S(E_{0},U_{0})=S(U_{esc},U_{esc})$ yields the original energy, $E_{0}$, of an atom escaping at a potential depth of $U_{esc}$. Repeating this measurement for many atoms, we partly reconstruct the atomic energy distribution. Figure \ref{Figure2} b shows the result for atoms that were trapped for 10 ms with the feedback enabled and disabled immediately before the potential was ramped down. It is clear from the figure that with feedback fewer atoms have a high energy, indicating a lower temperature. The low-energy part of the distribution was not observed because a further reduction of the final potential made our cavity lock unstable. 

For a more quantitative analysis we fit the Boltzmann distribution of a 3D harmonic oscillator, $p(E)dE=N\frac{E^{2}}{2(k_{B}T)^{3}}e^{-\frac{E}{k_{B}T}}dE$, to the data. As only atoms with an energy smaller than the trap depth $U_{0}$ are trapped, the energy distribution that we measure is normalized such that $\int^{U_{0}}_{0}p(E)dE=1$. The only fit parameter is the temperature $T$, which is found to be 160$\pm$5 $\mu$K with the feedback enabled and 400$\pm$50 $\mu$K without feedback where the errors are only due to the fit uncertainty. The observed temperature is comparable to the radial temperature of $T \leq 200 \mu K$ that has been observed using optical molasses cooling in a state-insensitive trap \cite{Boozer06, Boca04} inside a high-finesse cavity. Note that our temperature measurements suffer from a few systematic uncertainties. First, the assumed distribution is strictly valid only in a harmonic potential. But without feedback, the high temperature brings the atom into the anharmonic part of the potential. Second, heating effects during the ramp are neglected. Third, thermalization in three dimensions is assumed. However, this is not obvious as the cooling forces in the radial and axial directions are quite different. A detailed discussion of these issues is beyond the scope of this Letter.

The observed temperature is most likely limited by the low signal obtained from a well localized atom. Assuming an integration time of 13 $\mu$s and a probe power of 0.1 photons in the empty cavity, we can estimate the signal-to-noise ratio (shot noise of the weak probe signal) as a function of the radial atomic position. We find a value of 1 for an atom at a position that corresponds to an energy of about 450 $\mu$K. As the average energy of a particle in a 3D harmonic potential is $\bar{E}=3k_{B}T$, this simple signal-to-noise estimation predicts a temperature limit on the order of 150 $\mu$K, which is close to the observed temperature. Note that the relationship between cavity transmission and radial position is highly nonlinear. For example, even for a transmission signal twice as large the temperature limit would only drop to 135 $\mu$K. A significant temperature reduction could become feasible by tuning the probe laser to one of the normal modes where the transmission for well localized atoms is much higher. 

\begin{figure}
\includegraphics[width=8cm]{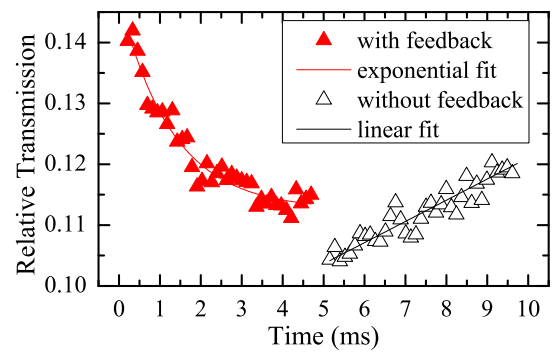}
\caption{\label{Figure3}(Color online) Averaged time-dependent transmission (relative to empty cavity) in two successive intervals in which the feedback is enabled ($\blacktriangle$) and disabled ($\vartriangle$), respectively. When the feedback is enabled, an exponential drop in transmission with a 1/e time constant of 1.2 ms is observed proving cooling of the atomic motion. Without feedback the average transmission increases linearly due to heating.}
\end{figure}

To show how feedback cooling improves the atomic localization, we trap an atom and enable and disable the feedback iteratively for 5 ms, switching to the deeper trapping potential of 950 $\mu$K when the feedback is disabled. The time-dependent transmission of the probe light in each of the two proper intervals is averaged over many intervals from a few hundred individually trapped atoms. The result is shown in Fig. \ref{Figure3}. To exclude intervals where the atom is leaving, we require the transmission during the two intervals and the following interval to be less than 50 $\%$ of the empty cavity transmission. As soon as the feedback is enabled, a sharp drop of the transmission is observed which approaches a steady state after a few ms. Fitting an exponential yields a time constant of 1.2 ms. This proves that feedback does indeed reduce the average radial oscillation amplitude. From the temperature measurement we deduce an average radial excursion of about 5 $\mu$m compared to about 8 $\mu$m without feedback cooling. We also see a linear increase in the transmission when the feedback is disabled. This is attributed to probe-light induced radial heating. The offset between the two curves comes from the change of the light shift caused by the change of the trap depth. For a shallower trapping potential, the atom-cavity detuning is larger which increases the transmission on the cavity resonance.

\begin{figure}
\includegraphics[width=8cm]{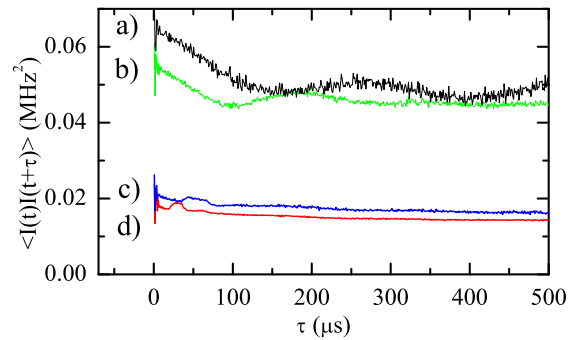} 
\caption{\label{Figure4}(Color online) Time-dependent intensity-correlation function without feedback in the shallow (a, black) and deep (b, green) trap, and with feedback using an integration time of the feedback algorithm of 32 $\mu$s (c, blue) and 16 $\mu$s (d, red). Without feedback the radial oscillations cause a characteristic bump at one-half of the oscillation period. This bump disappears when the feedback is enabled.}
\end{figure}

A complementary way to characterize feedback cooling is to look at the intensity-correlation function $\left\langle I(t)I(t+\tau)\right\rangle_{t}$ of the probe light as shown in Fig. \ref{Figure4}. To obtain these graphs we divide the experimental traces in intervals with duration $T$ = 2 ms. For intervals with transmission below 80\% of the empty cavity transmission we then evaluate $R(\tau)=\frac{1}{T-\tau}\int^{T-\tau}_{0}I(t)I(t+\tau)dt$ where $I(t)$ is the photon count rate averaged over 1 $\mu$s. We average over many intervals. The upper two curves show the average $R(\tau)$ for an atom trapped in the deep and the low trapping potential, respectively, when the feedback is disabled. Both curves have a characteristic bump at $T_{1}$ = 190 $\mu$s for the deeper potential and at $T_{2}$ = 270 $\mu$s for the shallower potential. This bump comes from the radial oscillations. $R(\tau)$ of atoms with enabled feedback is plotted in the lower part of the figure for integration times of the feedback algorithm of 16 $\mu$s and 32 $\mu$s. Three points are noteworthy. First, the average transmission is reduced which directly shows the better localization. Second, the radial oscillation bump is almost completely suppressed, providing further evidence for a reduction of the radial oscillation amplitude due to feedback cooling. Third, an almost square-shaped feature is observed for short correlation times. Its rising edge and width are given by the integration time of the feedback algorithm. This feature is an artifact of our feedback algorithm that preferentially switches the dipole trap with a period given by the integration time. As mentioned before, the change in the light shift has an effect on the transmitted signal, which is therefore modulated at the same frequency. 

The achieved trapping times put feedback cooling on par with state-of-the-art optical cooling in cavity QED experiments. We now use it routinely to enhance the experimental duty cycle and to improve the atomic localization. A more sophisticated tracking technique \cite{Horak02} or feedback algorithm can lead to further improvements. It will be interesting to see if it can be employed to cool the much faster axial motion along the standing-wave cavity axis. For strong confinement it might even be possible to cool the atom into the motional ground state \cite{Steck04}. As feedback cooling, in principle, does not rely on spontaneous emission of photons, it is particularly interesting for applications that require preservation of the internal state of the atom. Moreover, one can speculate about the possibility to (stochastically) cool ensembles of trapped atoms or molecules by means of feedback \cite{Raizen98,Balykin01,Ivanov03,Vilensky09}.

We thank J. Almer, B. Hagemann, and I. Schuster for early contributions to the experimental setup, J. Bayerl for technical assistance, and K. Murr and T. Wilk for stimulating discussions and comments on the manuscript. Financial support from the Deutsche Forschungsgemeinschaft (Research Unit 635), the European Union (IST project AQUTE) and the Bavarian PhD program of excellence (QCCC) is gratefully acknowledged.

\end{document}